
\magnification=1200
\hsize=15 truecm
\vsize=22 truecm
\baselineskip 20 truept
\voffset=1.5 truecm
\parindent=1cm
\overfullrule=0pt
\def\Dbar{\hbox{{\it D}\hskip-4.5pt/}}
\def\Nbar{\hbox{{$\nabla$}\hskip-6.5pt/}}

\def\vsp{\vskip 0.4truecm \par}
\def\ts{\thinspace}

\def\ni{\noindent}

\def\rp{\item{}}
\def \r#1{[\ts {#1}\ts \ts ]} \def\rf#1{\item{\r{#1} \ }}

\centerline{\bf AXIAL ANOMALY IN THE PRESENCE OF}
\centerline{\bf THE AHARONOV--BOHM GAUGE FIELD }

\vskip 1truecm

\centerline{{\bf  Paola Giacconi $^*$, St\'ephane  Ouvry $^{**}$,
Roberto  Soldati $^*$ }
\footnote{$^{(1)}$}{\rm E--mail :
giacconi, soldati@infn.bologna.it, ouvry@ipncls.in2p3.fr,}   }

\vskip 1truecm

\centerline{ $^*$ Dipartimento di Fisica  ``A. Righi" Bologna,
sezione I.N.F.N. Bologna, Italia }

\centerline{$^{**}$ Division de Physique Theorique
\footnote{$^{(2)}$} {Unit\'e de Recherche des Universit\'es Paris XI et Paris
VI
                     associ\'ee au CNRS}  }
\centerline{ IPN.F-91406 Orsay, France and}
\centerline{ LPTPE, Universit\`e Pierre et Marie Curie F-75252 Paris}

\vskip 2.0 truecm

ABSTRACT:\quad We investigate  on the plane the axial anomaly for euclidean
Dirac fermions in the presence of a background Aharonov--Bohm gauge potential.
The non perturbative analysis depends on the self--adjoint extensions
of the Dirac operator and the result is shown to be influenced by the
actual way of understanding the local axial current.
The role of the quantum mechanical parameters involved in the expression
for the axial anomaly is discussed. A derivation of the effective action by
means of the stereographic projection is also considered.

\vfill\eject

\noindent
{\bf 1.\quad Introduction}
\bigskip

The behaviour of matter fields in the presence of Aharonov-Bohm (AB), or anyon,
gauge fields has recently attracted a lot of interest owing to its possible
application to condensed matter physics. Clearly,
the origin of the much studied bidimensional anyon model [1] is
intimately connected with topological AB interactions.
{}From the field theoretical
point of view, the coupling of 2-dimensional fermions with such a kind
of gauge fields exhibits some interesting features which are worthwhile to be
carefully investigated.

In this note we shall deal with the problem of
computing the axial anomaly [2], as well as the effective action, for an
euclidean 2-dimensional Dirac field interacting with a background AB
potential, whose flux intensity is given by $\alpha$, with $ -1<\alpha<0 $.
Besides its own theoretical interest, in Ref.[3]
the axial anomaly has
been shown  to be related to the second virial
coefficient of an anyon gas, thereby giving a connection with in
principle measurable
thermodynamical properties of anyonic matter.
Moreover it has also been observed that the 1+1 axial anomaly is connected
with measurable effect in solid state physics [4].
We shall more precisely concentrate on the dependence of the axial
anomaly on boundary conditions at the location of the flux tube, i.e. on
possible self--adjoint extensions of the Dirac hamiltonian. The result
of our analysis will be fully exhibited, on the plane, for the special value
$\alpha=-{1\over 2}$, owing to merely technical limitations.
As a matter of fact, it should be hopefully clear
that there is no reason,
in principle, to doubt that our main statements still hold true for the whole
range of $\alpha$.

We first treat the model in the 2-dimensional plane and later on we shall
study its compactification by means of the stereographic projection.

The starting point is the classical euclidean action,

$$
S = \int  d^2x  \bar{\psi}(x)(i\Dbar)\psi(x) \quad ,
\eqno{(1.1)}
$$

\noindent
where

$$
i\Dbar=i\gamma_{\mu} (\partial_{\mu} - ieA_{\mu})\quad,
\eqno{(1.2)}
$$
\noindent
the AB gauge potential being

$$
A_{\mu} \equiv {\alpha\over e} \epsilon_{\mu\nu} \partial_{\nu}
        \ln \sqrt{x_{1}^{2}+x_{2}^{2}}
        = {\alpha \over e} \epsilon_{\mu\nu} {x_{\nu}\over x_{1}^{2}+x_{2}^{2}}
\quad,
\eqno{(1.3)}
$$
\noindent
Here, $\gamma_1=\sigma_1,\ \ \gamma_2=\sigma_2,\ \ \gamma_3=i\gamma_1\gamma_2=
\sigma_3$, where the {\it sigma}'s  are the Pauli matrices.
As it is well known the field strength is

$$
F_{\mu\nu}= {1\over2} \epsilon_{\mu\nu} F,\ \ \ \
F = -{4\pi\alpha \over e}  \delta^{(2)}(x) \quad,
\eqno{(1.4)}
$$
\noindent
which corresponds to  an infinitely thin solenoid at the
origin; $\psi$ and $\bar{\psi}$ are independent euclidean spinors.\par
The basic tool which allows for the non perturbative definition of the axial
current and anomaly is the complex power of the Dirac
operator. In general,  complex powers of pseudo-elliptic invertible
operator on compact manifolds whithout boundary do indeed exist under
very general hypotheses [5]. Since, however, we are on the
whole 2-dimensional plane and in the presence of field strengths
with $\delta$-like
singularities, the only way available to set up the complex powers is by means
of the spectral theorem. To this aim let us consider the eigenvalues
and eigenfunctions of the Dirac operator . The crucial point is that,
in the presence of an AB gauge potential,
 a symmetric Dirac hamiltonian is selected if one considers
a domain spanned by spinors which are regular at
the origin, but then completeness of the eigenstate basis is lost.
In order to find a
{\underbar {complete}}
orthonormal basis which diagonalizes the Dirac operator it is necessary
to consider
its self-adjoint extensions [6][7].

To this concern let us choose polar coordinates $(r,\phi)$ on the plane; the
eigenvalue problem becomes

$$
\eqalign{
(\gamma_1[\cos \phi {\partial \over \partial r}-{1\over r} \sin \phi
 ({\partial \over \partial\phi}+i\alpha)] &+
 \gamma_2 [\sin \phi {\partial \over \partial r}+{1\over r} \cos \phi
 ({\partial \over \partial\phi}+i\alpha)]) \psi_{\lambda}(r,\phi) \cr
&= - i\mu\lambda\ \psi_{\lambda}(r,\phi)\quad , \cr}
\eqno(1.5)
$$
\noindent
where $\mu$ is a suitable mass parameter to fix the scale of the eigenvalues.
If we rescale the spinor wave functions as
$$
{1\over \sqrt \mu}\psi_\lambda(r,\phi)\longmapsto
\psi_{\lambda,n}(\xi,\phi)\equiv
\left|\matrix{\psi_\lambda^{(L)}(\xi)e^{in\phi}\cr
              \psi_\lambda^{(R)}(\xi)e^{i(n+1)\phi} \cr}\right|\quad,
\eqno{(1.6)}
$$
\noindent
with $n\in {\bf Z},\ \ \lambda \in {\bf R}-\{0\}$ and $\xi=\mu r$,
we get the eigenspinors regular at the origin when $n\not=0$, namely
$$
\psi_{\lambda,\pm n}(\xi,\phi)= {\sqrt {|\lambda| \over 4\pi}}
\left|\matrix{(\pm i) J_{\pm\nu}(|\lambda|\xi)e^{\pm in\phi} \cr
              sgn(\lambda)J_{\pm (\nu+1)}(|\lambda|\xi)e^{i(1 \pm n)\phi}
\cr}\right|\quad;
\eqno(1.7)
$$
\noindent
here $n \in {\bf N}$, $J_\nu$ being the Bessel function of order
$\nu(\pm n) \equiv \pm n+\alpha $.

On the other hand, the partial waves corresponding to $\nu(0) \equiv \alpha $
can not be both regular at the origin unless completeness of the
eigenfunctions is lost [6]. Then one has to consider
the self-adjoint extensions  of the Dirac operator by means of
the standard
Von Neumann method of the deficiency indices [8]. This leads
to a one-parameter family $\Dbar_\omega, \ \omega \in {\bf R}$, whose
domain is given by ${\cal D}(\Dbar_\omega)=\{\psi + \beta(\psi^{(+)}+
e^{i\omega} \psi^{(-)})|\psi \in C^{(\infty)}[0,+\infty ]\cap
H^1_2([0,{\infty}]),
\beta \in {\bf C}, \omega \in {\bf R}\}$, where
$$
\psi^{(\pm)}\equiv {\sqrt {\mu\over {\cal N}}}
\left|\matrix{ K_\alpha(\mu\xi)\cr
               \pm K_{(1+\alpha)}(\mu\xi)e^{i\phi}\cr}\right|\quad,
\eqno{(1.8)}
$$
\noindent
$K_\nu$ being the Basset--McDonald  function,
${\cal N}$ a normalization constant
and $H^1_2$ the Sobolev space
$$
H^1_2 ([0,{\infty}])=\{\psi |  \int_0^{2\pi}d\phi \int_0^{\infty} \xi d \xi\
\psi^{\dag} (\xi,\phi) \psi(\xi,\phi) <\infty \}\quad .
\eqno{(1.9)}
$$
\noindent
The corresponding eigenfunctions for $ \nu=\alpha$ can be written in the form

$$
\eqalign
{&\psi_{\lambda,0}^{(\omega)}(\xi,\phi)=
{\sqrt {|\lambda|\over 4\pi(1+\sin\theta(|\lambda|)\cos\nu\pi)}}\ \times\cr
&\left|\matrix{i\cos {\theta(|\lambda|) \over 2} J_\alpha (|\lambda|\xi)-
              i\sin {\theta(|\lambda|) \over 2} J_{-\alpha}(|\lambda|\xi) \cr
 sgn(\lambda) \left[\cos {\theta(|\lambda|) \over 2}
J_{(1+\alpha)}(|\lambda|\xi) +
               \sin {\theta(|\lambda|) \over 2}
J_{-(1+\alpha)}(|\lambda|\xi)\right]
               e^{i\phi}\cr}\right|\cr}\quad,
\eqno(1.10)
$$
where
$$
\tan\theta(|\lambda|) = |\lambda|^{2\alpha+1}\tan\omega\quad .
\eqno(1.11)
$$

\noindent
We would like to notice that
the eigenfunctions in eq.s~(1.7),(1.10) are improper eigenfunctions, since they
belong to eigenvalues of the continuous spectrum. They are suitably
normalized according to theory of the distributions, $viz.$

$$
\lim_{R\to\infty}\int_0^{\mu R}\xi d\xi\int_0^{2\pi}d\phi\
\psi^{\dag}_{n_1}(|\lambda_1|\xi,\phi)\psi_{n_2}(|\lambda_2|\xi,\phi)=
\delta_{n_1 n_2}\delta(\lambda_1-\lambda_2)\quad .
\eqno(1.12)
$$

Moreover, in order to obtain the correct normalization as in eq.~(1.12),
one has to put the contribution at the origin equal to zero,
thereby finding the relationship of eq.~(1.11). It should be stressed
that, rather than the requirement on the domain ${\cal D}(\Dbar_\omega)$,
which actually involves normalizable states, it is just the condition (1.12)
of being
a complete orthonormal family of improper eigenfunctions, leading eventually
to eq.s~(1.10),(1.11) (for the angular momentum component $n=0$). As a matter
of fact, the property of having a complete orthonormal basis turns out to be
necessary and sufficient for an operator to be
self--adjoint. Furthermore,
we see that, for any value of $\omega$,
the  spectrum is purely continuous and is provided by
the whole real line, but the point $\lambda=0$, the zero modes being absent
since they are not orthonormalizable.
\medskip
\noindent
{\bf 2.\quad The axial anomaly on the plane}
\bigskip

Once the eigenvalue problem has been solved, we are able
to set up the complex power by means of the spectral theorem.
The complex power of the dimensionless operator
$I_{\omega}^{-s}\equiv \left({\Dbar_\omega \over \mu}\right)^{-s}$
is defined by the  kernel
$$
\eqalign{
& <\xi_1,\phi_1|I_\omega^{-s}|\xi_2, \phi_2> \equiv
  K_{-s}(I_\omega;\xi_1,\phi_1,\xi_2,\phi_2)\cr
& = 2\ \sum_{n=1}^{\infty} \left [\int_0^{\infty} d\lambda\
\lambda^{-s} \psi_{\lambda,n}(\xi_1,\phi_1)
\psi_{\lambda,n}^{\dag}(\xi_2,\phi_2) + (n \rightarrow -n)\right ]\cr
& +2\ \int_0^{\infty} d\lambda
\ \lambda^{-s} \psi_{\lambda,0}^{(\omega)}
(\xi_1,\phi_1)\psi_{\lambda,0}^{(\omega)\dag}(\xi_2,\phi_2)\quad , \cr}
\eqno{(2.1)}
$$
which can be analytically extended to a meromorphic function of the complex
variable $s$; the key property is that the kernel of the complex power is
regular at $s=0$.
In particular, the value of its trace over spinor
indices, on the diagonal $(\xi_1,\phi_1)=(\xi_2,\phi_2)$,
can be explicitely evaluated either in the case
$\omega=0,\pi,\quad -1<\alpha<0$ or in the case $\alpha= -{1\over 2}, \quad
\omega\in {\bf R}$. Actually, to the aim of computing the anomaly,
 it is more useful to write down the explicit
form of the traces of the axial kernels, namely
$$
\eqalign{
&tr[\gamma_3 K_s(I_{\omega=0,\pi};\xi,\xi)]=
\Sigma_{-s}(\alpha,\xi)+\Sigma_{-s}(-\alpha,\xi)\cr
&+{\mu\over 2\pi}\xi^{s-2}{\Gamma({s+1\over 2})\Gamma(\alpha+1-{s\over 2})\over
\sqrt\pi(s-1)\Gamma({s\over 2})\Gamma(\alpha+{s\over 2})}+(\alpha\longmapsto
\alpha+1)\quad ,\cr}
\eqno{(2.2)}
$$
where we have set
$$
\Sigma_{-s}(\alpha,\xi)={\mu\over 2\pi}\xi^{s-2}
{\Gamma({s+1\over 2})\Gamma(\alpha+2-{s\over 2})\over
\sqrt\pi(s-1)(s-2)\Gamma({s\over 2})\Gamma(\alpha+{s\over 2})}\quad ,
\eqno(2.3)
$$
whereas
$$
\eqalign{
&tr[\gamma_3 K_s(I_{\omega};\xi,\xi)]|_{\alpha=-{1\over 2}}=\cr
&{\mu\over \pi\sqrt\pi}\xi^{s-2}{\Gamma({1-s\over 2})\over \Gamma({s\over 2})}
\left(1+\sin\omega{\Gamma({s\over 2})\over \Gamma({s+1\over 2})}\right)
\quad .\cr}
\eqno(2.4)
$$

Now we are ready to obtain the local forms of the axial anomaly, as it arises
from the non perturbative definition of the fermionic axial current
associated to the invertible
operators $I_{\omega}$.
As a matter of fact we notice that, since the
kernel $K_{-s}(I_{\omega};\xi,\xi)$
of the complex power is a well defined tempered distribution, depending
meromorphically upon the complex variable $s$, one can properly define
the euclidean averages of the vector and axial currents, respectively,
by means of point--splitting as well as analytic continuation [9], namely
$$
\eqalign{
&<j_\mu^{(\omega)}(x)>=e<tr[\gamma_\mu\psi(x)\psi^{\dag}(x)]>\equiv\cr
&\lim_{s\to 1}\lim_{\epsilon\to 0}e\
tr[\gamma_\mu K_{-s}(I_\omega;x,x+\epsilon)]
=\cr
&\lim_{s\to 1}\lim_{\epsilon\to 0}\sum_{n=-\infty}^{+\infty}\int_{-\infty}^{+
\infty}d\lambda\ \lambda^{-s}\ e\ tr[\gamma_\mu\psi_{n,\lambda}^{(\omega)}(x)
\psi_{n,\lambda}^{(\omega)\dag}(x+\epsilon)]\quad ,\cr}
\eqno (2.5a)
$$
where $<\cdot>$ means euclidean average and
$$
\eqalign{
&<j_{\mu 3}^{(\omega)}(x)>=e<tr[\gamma_\mu\gamma_3\psi(x)\psi^{\dag}(x)]>
\equiv\cr
&\lim_{s\to 1}\lim_{\epsilon\to 0}e\ tr[\gamma_\mu\gamma_3
K_{-s}(I_\omega;x,x+\epsilon)]
=\cr
&\lim_{s\to 1}\lim_{\epsilon\to 0}
\sum_{n=-\infty}^{+\infty}\int_{-\infty}^{+
\infty}d\lambda\ \lambda^{-s}\ e\ tr[\gamma_\mu\gamma_3
\psi_{n,\lambda}^{(\omega)}(x)
\psi_{n,\lambda}^{(\omega)\dag}(x+\epsilon)]\quad .\cr}
\eqno (2.5b)
$$
{}From the above definitions of the averaged local currents
it is straightforward to show, taking eq.~(1.5) into
account, the quantum balance equations, namely
$$
<\partial_\mu j_\mu^{(\omega)}(x)> = 0
\quad ,
\eqno (2.6a)
$$
testing the gauge invariance of the definition in eq.~(2.5a), whereas
$$
<\partial_\mu j_{\mu 3}^{(\omega)}(x)> = 2ie\ \lim_{s\to 0}
\ tr[\gamma_3\
K_s(I_\omega;x,x)]\equiv {\cal A}^{(\omega)}(x)\quad
\eqno (2.6b)
$$
leads to the definition of the local axial anomaly, once the topology has
been chosen in taking the limit $s\to 0$; we shall discuss below this delicate
matter. \par

First we consider the ${\cal S}^\prime$-topology, namely we study
$$
\lim_{s\to 0}\ \int d^2x\ tr[\gamma_3\ K_s(I_\omega;x,x)]\ f(x)\quad ,
$$
for any rapidly decreasing function
$f\in {\cal S}({\bf R}^2)$.

If we take the limit $s\to 0$ in the sense of the distributions, from
eq.~(2.2) we get
$$
\int d^2 x\  {\cal A}^{(0,\pi)}(x)f(x)= -\alpha\quad ,
\eqno(2.7)
$$
where $f$ is a suitable test function belonging to ${\cal S}({\bf R}^2)$
normalized to $f(0)=\mu$; we notice that the above result, in
full agreement with the one of Ref.[3], actually corresponds to the usual
result, $viz.$ ${\cal A}^{(0,\pi)}(x)=-{ie^2\over 2\pi}\epsilon_{\mu\nu}
F_{\mu\nu}(x)$ as a distribution.

On the other hand, we can start from eq.~(2.4), with $\alpha=-{1\over 2}$,
and try to take the limit $s\to 0$, for any $\omega$, in the
${\cal S}^\prime$-topology. Let us consider indeed in the RHS of eq.~(2.4)
the term proportional to $\sin\omega$. Then from the identity
$$
\eqalign{
&{\cal S}^\prime-\lim_{s\to 0}\ \left(\xi^{s-2} -{2\pi\over s}\delta^{(2)}(\xi)
\right)=\cr
&{1\over 2\xi}{d\over d\xi}(\xi{d(\ln\xi)^2\over d\xi})\quad ,\cr}
\eqno(2.8)
$$
it immediately follows that the ${\cal S}^\prime$-limit does not exist unless
$\omega=0,\pi$ and, in this case, eq.~(2.7) still holds with
$\alpha=-{1\over 2}$.
The above analysis strongly suggests that the definition of the anomaly, as
the limit in the sense of distributions of the axial kernel, leads to select
$\omega= 0,\pi$ as the correct physical choices for the self--adjoint
extensions
of the Dirac operator. The same conclusion has been claimed in the literature
[7] from a quite different point of view, namely
by demanding that one does not have any additional contact interaction at
the origin beyond the point--like magnetic field.
Nevertheless it is worthwhile to mention that some {\it dual} point of view
leads to a different outcome, when we consider the limit $s\to 0$ in the
natural
topology of ${\bf R}-\{0\}$ and {\underbar {only afterwards}}
continue the result to
${\cal S}^\prime ({\bf R}^2)$. As a matter of fact, a non vanishing result is
obtained in this case for $\omega\not= 0,\pi$ and, when $\alpha=-{1\over 2}$,
we
can compute explicitely
$$
\lim_{s\to 0}\lim_{\epsilon\to 0}e\ tr[\gamma_3
K_{-s}(I_\omega;x,x+\epsilon)]|_{\alpha=-{1\over 2}}
={ie\sin\omega\over 2\pi^2r^2},\quad r\not= 0\quad .
\eqno(2.9)
$$
As a consequence, there is a unique continuation in ${\cal S}^\prime({\bf
R}^2)$
which reads ($\alpha=-{1\over 2}$)
$$
\int_0^{\infty} \xi d\xi \int_0^{2\pi}d{\phi}\ f(\xi,\phi){\cal A}^{(\omega)}
(\xi)= \int_0^{\infty} \xi d\xi \int_0^{2\pi}d{\phi}\ {ie\sin\omega\over
2\pi^2[\xi^2]}\quad ,
\eqno(2.10)
$$
where $f$ is a test function belonging to ${\cal S}({\bf R}^2)$.
We recall the definition of the tempered distribution
$$
{1 \over {[\xi^2]}} \equiv {1 \over 2\mu^2}(\partial_1^2 + \partial_2^2)
                      (\ln r)^2 + C(\mu) \delta^{(2)} (x)\quad ,
\eqno(2.11)
$$
$C(\mu)$ being an {\underbar {arbitrary}} function of the regularization mass
parameter. It should be stressed that the freedom (actually up to $\ln \mu$)
within $C(\mu)$ amounts to the natural requirement
$$
{1 \over {[\xi^2]}}= {1\over \mu^2}\cdot{1\over [r^2]}\quad ;
\eqno (2.12)
$$
consequently we can think about $C(\mu)$ in terms of a scaling function,
in the sense that, when $\mu\to p\mu$, we have $C(\mu)\to C(p\mu)=
C(\mu)+2\pi\ln p$.\par

Now some remarks are in place to comment this result. On the one hand,
we observe that
eq.~(2.10) involves two further quantum mechanical arbitrary parameters,
$\omega$ and the scaling factor $C(\mu)$, beyond the flux intensity
$\alpha=-{1\over 2}$ of the classical background infinitely thin solenoid.
In particular,
 if the test function vanishes at the origin, where
 the field strength is concentrated, still a nonvanishing contribution
survives, of a purely quantum mechanical nature, which depends upon the
parameter of the self--adjoint extensions.
On the other hand, we recall that in the case of smooth gauge fields
the local
axial anomaly turns out to be proportional to the field strength itself, up
to the presence of zero modes (see Refs. [3][9][10] and the last part of the
present paper).
In the presence of the singular AB gauge potential,
the local axial anomaly still exhibits the above feature if we define the axial
current as a limit in the ${\cal S}^\prime$--topology, and, moreover, the
self--adjoint extension is fixed.
Alternatively, the local axial anomaly
is different from zero even outside
the infinitely thin solenoid, and the parameter of the self--adjoint extensions
is free, if we define the axial current and anomaly as in eq.s~(2.9),(2.10).
We could say that this latter quite interesting feature
is closely reminescent of the AB effect.\par

In conclusion some physical, {\it a priori} measurable quantity [3][4],
such as the local axial anomaly,
in the present singular case, appears to be reproduced by the standard
Schwinger's [11] form
${\cal A}(x)=-{ie^2\over 2\pi}\epsilon_{\mu\nu}
F_{\mu\nu}(x)$, the parameter of the self--adjoint extensions being fixed
to $\omega = 0,\pi$, if the regularized axial current is defined as a limit
in the topology of tempered distributions.
Nonetheless, from the mathematical point of view, some inequivalent
construction of the singular axial current might be considered, which
eventually leads to a non standard form of the axial anomaly and to the
freedom in the choice of the self--adjoint extensions. Strictly
speaking, the above statements holds true in the special cases we have
explicitely worked out. Nevertheless it is highly presumable that the same
features still appear in general, although the explicit proof does not
seem to be presently available.
The physical content of this last approach will be discussed elsewhere.
\medskip
\noindent
{\bf 3.\quad The effective action}
\bigskip
Another important issue is the calculation of the effective action. To this
purpose it should be noticed that the standard gauge invariant Schwinger
formula on the plane [11] does not make sense,
owing to the singular nature
of the AB gauge potential. Nevertheless, a meaningful definition of the
effective action may be obtained [9][12], through the transition to the
compact case
by means of the stereographic map from the plane to the open 2--sphere
(without the north--pole),
the origin of the plane coinciding with the south--pole.
The open 2--sphere is embedded in ${\bf R}^3$ in such a way that a convenient
choice of the coordinates reads
$$
\eqalign{
&X_1=2a\ {y\cos\phi\over 1+y^2}\quad , \cr
&X_2=2a\ { y\sin\phi\over 1+y^2}\quad , \cr
&X_3=a\ {1-y^2\over 1+y^2}\quad ,\cr
&X_{\mu}X_{\mu}+X_3^2=a^2,\quad  y={r\over a}\quad .\cr}
\eqno(3.1)
$$
It is immediate to obtain the form of the zwei--beine,
namely
$$
\eqalign{
&e_{a\mu}(y)={2\over 1+y^2}\delta_{a\mu}\quad ,  \cr
&E_{a}^\mu(y)={1+y^2\over 2}\delta_{a\mu}\quad .\cr}
\eqno(3.2)
$$
The invariant measure becomes
$$
\int d\mu\equiv \int_0^{2\pi} d\phi\int_0^{\infty}ydy {4a^2\over
(1+y^2)^2}\quad .
\eqno(3.3)
$$
Within the above conformal coordinate system ${x_\mu}$,
the eigenvalue problem for the covariant
Dirac operator takes the form
$$
{1+y^2\over 2}\gamma_\mu\left(\nabla_\mu-{1\over 2}\partial_\mu \ln{1+y^2\over
2}\right)\psi_\lambda (x)+i{\lambda\over a}
\psi_\lambda (x)=0\quad ,
\eqno(3.4)
$$
where
$$
\nabla_\mu\equiv \partial_\mu -ieA_\mu(x)\quad .
\eqno(3.5)
$$
If we pass to polar coordinates and perform a conformal mapping on the spinors,
namely if we set
$$
\sqrt a\ \psi_{\lambda,n}(x)\longmapsto \sqrt{{2\over 1+y^2}}
\left|\matrix{\psi_\lambda^{(L)}(y)e^{in\phi}\cr
              \psi_\lambda^{(R)}(y)e^{i(n+1)\phi} \cr}\right|\quad,
\eqno{(3.6)}
$$
the eigenvalues problem becomes equivalent to the system of
coupled differential equations
$$
\eqalignno{
&y{d\psi_\lambda^{(L)}\over dy}-\nu\psi_\lambda^{(L)}=
{2\lambda y\over i(1+y^2)}
\psi_\lambda^{(R)} &(3.7a) \cr
&y{d\psi_\lambda^{(R)}\over dy}+(\nu+1)\psi_\lambda^{(R)}=
{2\lambda y\over i(1+y^2)}
\psi_\lambda^{(L)}, &(3.7b)\cr}
$$
where $\nu(\pm n)=\pm n+\alpha$, $\nu(0)\equiv \alpha$.
In order to solve eqs.~(3.7) it turns out to be useful to perform the
change of
variables
$$
\rho\equiv {y^2-1\over y^2+1},\quad \int d\hat\mu(\rho,\phi)=\int_0^{2\pi}
d\phi
\int_{-1}^1{2 d\rho\over 1-\rho};
\eqno(3.8)
$$
Hence, the eigenvalues problem takes the form
$$
\eqalignno{
&(1-\rho^2){d\chi^{(L)}_l\over d\rho} - \nu \chi^{(L)}_l + i\lambda_l\sqrt{1-
\rho^2}
\chi^{(R)}_l=0\quad , &(3.9a)\cr
&(1-\rho^2){d\chi^{(R)}_l\over d\rho} + (\nu+1) \chi^{(R)}_l +
i\lambda_l\sqrt{1-\rho^2}
\chi^{(L)}_l=0\quad , &(3.9b)\cr}
$$
with $\lambda_l\not=0$ and
$$
\eqalignno{
&(1-\rho^2){d\chi^{(L)}_0\over d\rho} - \nu \chi^{(L)}_0 =0\quad , &(3.10a)\cr
&(1-\rho^2){d\chi^{(R)}_0\over d\rho} +
(\nu+1) \chi^{(R)}_0 =0\quad , &(3.10b)\cr}
$$
for the zero modes.\par
We would like to notice that the covariant Dirac operator on the open
2--sphere,
up to the conformal mapping (3.6),
turns out to be essentially self--adjoint with domain given by square
integrable functions on the interval $\rho\in [-1,1]$, the derivatives at
$\rho=\pm 1$
being understood in the sense of the distributions. In order to fulfil
the above requirement, only the value $n=0$ is allowed, a feature which leads
to the removal of the degeneracy with respect to the orbital angular momentum.
The complete set of orthonormal eigenfunctions reads
$$
\eqalignno{
&\chi^{(L)}_l(\rho,\phi)=
{1\over 2\sqrt{\pi}}\left({1+\rho\over 1-\rho}\right)^{{\alpha\over
2}}{\hat P}_l^{(-\alpha - 1, \alpha)}(\rho)\quad , &(3.11a)\cr
&\chi^{(R)}_l(\rho,\phi)= {\pm i e^{i\phi}\over 2\sqrt{\pi}}
(1-\rho)\left({1+\rho\over 1-\rho}\right)^{{1+\alpha\over
2}}{\hat P}_{l-1}^{(-\alpha, 1+\alpha)}(\rho)\quad , &(3.11b)\cr}
$$
with eigenvalues $\lambda_l=l^2,\quad l=1,2,...$.  The
sign indetermination is related to the global euclidean axial symmetry
$\psi\longmapsto e^{\gamma_3 \eta}\psi$. Here ${\hat P}_l^{(\alpha,\beta)}
(\rho)$
stands for the normalized Jacobi polynomials [13], namely
$$
\eqalign{
&{\hat P}_l^{(\alpha,\beta)}(\rho)\equiv
\sqrt{{l!(2l+\alpha+\beta+1)\Gamma(l+\alpha
+\beta+1)\over 2^{\alpha+\beta+1}\Gamma(l+\alpha+1)\Gamma(l+\beta+1)}}
\ P_l^{(\alpha,\beta)}(\rho)\quad ,\cr
&P_l^{(\alpha,\beta)}(\rho)\equiv
{(-1)^l\over 2^ll!}(1-\rho)^{-\alpha}(1+\rho)^{-\beta}{d^l\over d\rho^l}
[(1-\rho)^{\alpha+l}(1+\rho)^{\beta+l}\quad .\cr}
\eqno(3.12)
$$
The noteworthy feature is the appearence of zero modes, namely
$$
\eqalignno{
&\chi^{(L)}_{0}(\rho)= {1\over 2\pi}\sqrt{\sin(-\pi\alpha)}
\left({1+\rho\over 1-\rho}\right)^{{\alpha\over 2}}\quad , &(3.13a)\cr
&\chi^{(R)}_{0}(\rho)=0 \quad , &(3.13b)\cr}
$$
which indicates the non trivial topological behaviour of the AB potential.
By the way,
the reason for the vanishing of the right--zero--mode comes from the
requirement of orthonormality for the complete set of the eigenfunctions of
the self--adjoint Dirac operator on the open 2--sphere.\par
Let us come to the explicit evaluation of the quantum effective action
and of the corresponding axial anomaly.
As it is known, the 2--dimensional classical action for the Dirac spinor field
on the open 2--sphere is conformally equivalent to the
corresponding quantity on the complex plane, namely [12]
$$
\eqalign{
&W_{plane}\equiv W_{sphere}+
{1\over 24\pi}\int d^2x\ \omega(x)\triangle\omega(x)\quad ,\cr
&\omega(x)=\ln {1+({x\over a})^2 \over 2}\quad ,\cr}
\eqno (3.14)
$$
where $W$ stands for the logarithm of the regularized determinant of the Dirac
operator.
As a consequence, taking eq.~(3.6)
into account, we can {\underbar {define}} the regularized determinant of the
flat Dirac operator by means of the zheta--regularization of the associated
self--adjoint operator of eqs.~(3.4), which has a discrete spectrum.
Actually, as the
spectrum includes the null eigenvalue, we have to set [9]
$$
\zeta (ia\hat{\Nbar} ; s )\equiv
\sum_{l=1}^\infty (l^2)^{-s}=\zeta (2s)\quad ,
\eqno(3.15)
$$
where
$$
\hat {\Nbar}\equiv {1+({x\over a})^2 \over 2} \Nbar_x \quad .
\eqno(3.16)
$$
and consequently
$$
W_{plane}= -\ln {\det}^{\prime}(ia\hat{\Nbar})\equiv
{d\over ds} \zeta (ia\hat{\Nbar}; s )|_{s=0} =
-{1\over 2}\ln 2\pi \quad .
\eqno(3.17)
$$
whereas
$$
W_{sphere}={1\over 6}(1-\ln 2)-{1\over 2}\ln 2\pi\quad .
\eqno (3.18)
$$
The local axial anomaly is now arising from the variation (see Refs. [8][14])
$$
\int d^2x\  {\cal A}(x)f(x) = 2ie
\int d^2x\ {\delta W_{plane}\over \delta A_\mu(x)}
\ {i\over e}\epsilon_{\lambda\mu}\partial_\lambda f(x)\quad ,
\eqno (3.19)
$$
and leads, nota bene, to a one--parameter family of local expressions, the
parameter being given by the compactification radius $a$, namely
$$
{\cal A}(x)= 2ie\lim_{s\to 0}\sum_{l=1}^{\infty}
(l^2)^{-s}\ tr[\gamma_3\psi_l({x\over a})\psi_l^\dagger({x\over a})]
\quad .
\eqno (3.20)
$$
Unfortunately, it does not seem that the RHS of the last equation could be
set into a close analytical form, just as in the general case of the continuous
spectrum on the plane.
It should be stressed
that the standard Seeley--DeWitt asymptotic expansion
can not be safely used to handle the RHS of eq.~(3.17),
since the original theory
is strictly defined on the open 2--sphere without poles; in
particular the fermion eigenfunctions are singular (not meromorphically)
at the poles. Nevertheless, one can try to approximate the present singular
case
by means of a smooth "vortex--like" gauge potential, namely one can consider
$$
A_{\mu}(x)=\  {\cal S}^\prime-\lim_{\sigma\to 0}\
{\alpha \over e} \epsilon_{\mu\nu} {x_{\nu}\over x_{1}^{2}+x_{2}^{2} +
\sigma^2}
\quad ,
\eqno(3.21)
$$
the limit being taken at the end of the anomaly calculation (actually this
corresponds to a regularization of the singular statistical AB interaction).
Within this framework, the anomaly can be computed from the standard methods
[9],[10] and, in the decompactification limit $a\to \infty$, it reproduces
the usual Schwinger's result, the topological zero modes being disappeared.
It should be noticed that this result is in agreement with what we have
previously found on the plane, if the anomaly is coherently understood as
a limit in the topology of tempered distributions (see eq.~(2.7)).
In this sense we argue
that the results on the plane, we have previously discussed,
should be of general character.
\par

In conclusion we find that the effective action
can be computed exactly (see eq.~(3.17)) starting from the
stereographic map, whereas
the local axial anomaly can be
explicitely evaluated on the plane, where the
effective action does not exist in the present singular case. Here we have
presented the detailed calculation of the local axial anomaly on the plane only
for particular values of the parameters.
Obviously, a technical effort should be attempted
in order to work out exactly the quite
general cases for the effective action as well as for the anomaly.
It would also be very interesting to investigate other kinds of physical
situations such as,
in particular,
the occurrence of confining potentials and/or smooth boundaries,
to further understand their influence on the axial anomaly and the effective
action.
Moreover we think to analyze the eventual relevance of eq.~(2.10) in the
evaluation of many body hamiltonian, virial coefficients and
other physical effects.

\bigskip
{\bf Acknowledgments}: \par\noindent
P.G. and R.S. warmly thank A. Bassetto, M. Mintchev
and G. Nardelli for helpful discussions. S.O. thanks A. Comtet, A. Khare and
J. McCabe for numerous discussions on the subject.
\bigskip

\vsp
\ni{{\bf  References}} \vsp
\rf1  J.M. Leinaas and J. Myrheim, Nuovo Cimento {\bf B37} (1977) 1;
F. Wilczek, "Fractional Statistics and Anyon Superconductivity", World
Pub. (1990); for a recent review see also J. Myrheim,
Unit Report Trondheim (September 1993).
\rf2  S.L. Adler, Phys. Rev. {\bf 177} (1969) 2426;
\rp   J. Bell and R. Jackiw, Nuovo Cimento A {\bf 60} (1969) 47.
\rf3  A. Comtet and S. Ouvry, Phys. Lett. {\bf 225B} (1989) 272.
\rf4  I.V. Krive and A.S. Rozhavsky, Phys. Lett. {\bf 113A} (1985) 313;
\rp   Zhao--bin Su and B. Sakita, Phys. Rev. Lett. {\bf 56} (1986) 780 and
      Prog. Theor. Phys. Supplement {\bf 86} (1986) 238.
\rf5  R.T. Seeley, Ann. Math. Soc. Proc. Symp. pure Math.
      {\bf 10} (1967) 288;
\rp   P.B. Gilkey, "{\it Invariance theory, the heat equation, and
      Atiyah-Singer index theorem}", Publish or Perish, Boston (1974).
\rf6  P. de Sousa Gerbert, Phys. Rev. {\bf D 40} (1989) 1346;
\rp   P. de Sousa Gerbert and R. Jackiw,
      Comm. Math. Phys. {\bf 124} (1989) 229;
\rp   R. Jackiw, in "{\it M.A.B. Beg Memorial Volume}", World Scientific Pub.,
      Singapore (1991).
\rf7  C. Manuel and R. Tarrach, Phys. Lett. {\bf 268B} (1991) 222 and Phys.
      Lett. {\bf 301B} (1993) 72;
\rp   J. Grundberg, T.H. Hansson, A. Karlhede and J. M. Leinaas,
      Mod. Phys. Lett. {\bf B} (1991) 539.
\rf8  M. Reed and B. Simon,
      "{\it Fourier Analysis and Self-Adjointness}",
      Academic Press, Orlando (1987);
\rp   for point--like interactions see
      S. Albeverio, F. Gesztesy, R. Hoegh--Krohn and H. Holden, "{\it
      Solvable Models in Quantum Mechanics}", Springer--Verlag, New York
      (1988).
\rf9  M. Hortacsu, K.D. Rothe and B. Schroer, Phys. Rev. {\bf D 20}
      (1979) 3023;
\rp   in the context of the Schwinger model see also F. Strocchi, "{\it A
      short introduction to the foundations of quantum field theory}",
      Pisa--Trieste lecture notes (1993).
\rf{10}  A. Bassetto and L. Griguolo, Jour. Math. Phys. {\bf 32} (1991) 3195.
\rf{11} J. Schwinger, Phys. Rev. {\bf 128} (1962) 2425; {\it Theoretical
      Physics}, Trieste Lectures (1962), p. 89, I.A.E.A. Vienna (1963).
\rf{12} J.S. Dowker, Manchester prep. MUPT/93/24 (1993) and references
      therein.

\rf{13}  I.S. Gradshteyn and I.M. Ryzhik, "{\it Table of integrals series
      and products}", Academic Press, San Diego (1979).
\rf{14}  A. Bassetto, L. Griguolo and R. Soldati, Phys. Rev. {\bf D 43} (1991)
         4088.

\vfill\eject
\bye